\documentclass[12pt]{article} \topmargin= 0.3cm \textwidth 16.5cm
=0cm \headsep =0cm
\usepackage {latexsym}
\usepackage {graphicx}

\begin{document}

\title{Second Randall-Sundrum brane world scenario with a nonminimally coupled bulk scalar field}
\author{K. Farakos \footnote{kfarakos@central.ntua.gr} and P. Pasipoularides  \footnote{paul@central.ntua.gr} \\
       Department of Physics, National Technical University of
       Athens \\ Zografou Campus, 157 80 Athens, Greece}
\date{ }
       \maketitle

\begin{abstract}
In our previous work of Ref. \cite{FP} we studied the stability of
the RS2-model with a nonminimally coupled bulk scalar field $\phi$,
and we found that in appropriate regions of $\xi$ the standard
RS2-vacuum becomes unstable. The question that arises is whether
there exist other new static stable solutions where the system can
relax. In this work, by solving numerically the Einstein equations
with the appropriate boundary conditions on the brane, we find that
depending on the value of the nonminimal coupling $\xi$, this model
possesses three classes of new static solutions with different
characteristics. We also examine what happens when the fine tuning
of the RS2-model is violated, and we obtain that these three classes
of solutions are preserved in appropriate regions of the parameter
space of the problem. The stability properties and possible physical
implications of these new solutions are discussed in the main part
of this paper. Especially in the case where $\xi=\xi_c$ ($\xi_c$ is
the five dimensional conformal coupling) and the fine tuning is
violated, we obtain a physically interesting static stable solution.
\end{abstract}

\section{Introduction}

A possible extension of conventional four-dimensional field theory
models is based on the well known brane world scenario. According to
this scenario, ordinary matter is assumed to be trapped in a
submanifold with three spatial dimensions (brane world) that is
embedded in a multi-dimensional manifold (bulk). Contrary to
ordinary matter, gravitons are allowed to propagate in the bulk. The
new feature in this scenario is that the extra dimensions can be
large or even infinite. In addition brane world models predict
interesting phenomenology even at TeV scale \cite{ADD}, and put on a
new basis fundamental problems such as the hierarchy and the
cosmological constant problem (see also the reviews \cite{Rub,Lon}
and references therein).

The typical example of a Brane-world model with an infinite warped
extra dimension is the second Randall-Sundrum model (RS2-model)
\cite{Ran}. In this scenario, we have a single brane with a positive
energy density (the tension $\sigma$), whereas the bulk has a
negative five dimensional cosmological constant $\Lambda$. The
corresponding Einstein equations have a solution only if a
fine-tuning condition is satisfied
($\Lambda=-\frac{\sigma^{2}}{6}$), in units where $8 \pi G_5=1$.
This solution implies a space-time geometry of the form of $AdS_{5}$
around the brane, and Minkowski with zero effective cosmological
constant on the brane. An extension of the RS2-model with a second
negative tension $-\sigma$ brane, is the RS1-model \cite{Ran}. In
this case we have an orbifolded extra dimension of radius $r_c$. The
two branes are fitted to the fixed points of the orbifold, $z=0$ and
$z_c=\pi r_c$ with tensions $\sigma$ and $-\sigma$ correspondingly.
The particles of the standard model are assumed to be trapped on the
negative tension brane, which is called visible, while the positive
tension brane is called hidden.

In this work we study the RS2-model with a nonminimally coupled bulk
scalar field, via an interaction term of the form $-\frac{1}{2}\xi R
\phi^2$, where $\xi$ is a dimensionless coupling \footnote{The
coupling $\xi$ possesses two characteristic values: a) the minimal
coupling for $\xi=0$ and b) the conformal coupling for
$\xi_{c}=3/16$ }. The motivation is to examine whether this model
possesses static solutions other than the standard one of the
RS2-model (see Eq. (2) below).

In particular we show, by solving numerically the Einstein equations
with the appropriate boundary conditions on the brane, that
according to the value of the nonminimal coupling $\xi$ our model
possesses three classes of new static solutions: (a) for $\xi<0$ the
solutions develop a naked singularity in the bulk, (b) for
$\xi>\xi_{c}$ (where $\xi_{c}$ is the five dimensional conformal
coupling) we find that the warp factor $a(z)$ is of the order of
unity near the brane and increases exponentially ($a(z)\sim e^{k
z}$), as $z\rightarrow +\infty$, while the scalar field $\phi(z)$ is
nonzero on the brane and tends rapidly to zero in the bulk (in this
case the space-time is asymptotically $AdS_{5}$), and (c) for
$0<\xi<\xi_{c}$ the warp factor $a(z)$ and the scalar field
$\phi(z)$ tend rapidly to infinity. Contrary to case (b), where the
scalar curvature is asymptotically constant, in case (c) the scalar
curvature tends to infinity. In addition we examine what happens
when the fine tuning of the RS2-model is violated, and we obtain
that in appropriate regions of the parameters of the problem the
three classes of solutions we described above are preserved.

This work is motivated by our previous paper \cite{FP}. In
particular in \cite{FP} we investigated the spectrum of a
nonminimally coupled bulk scalar field in the background of the
RS2-metric. We obtain that for $\xi<0$ the spectrum of the scalar
field exhibits a unique bound state with negative energy, or a
tachyon mode. Note that the existence of a localized tachyon mode
indicates a gravity-induced Dvali-Shifman Mechanism \cite{Shif} as
we argue in \cite{FP}. It is worth to note, that this mechanism for
the localization of Gauge field, from the point of view of lattice,
has been investigated in Refs.\cite{lat,Rumm}. In \cite{FP} we had
not examined the spectrum of the scalar field in the case of
$\xi>0$. We complete this investigation here. In the region $0 \leq
\xi \leq \xi_{c}$ we obtain that there is no tachyon mode (see
Appendix A). However the tachyon mode returns \footnote{In this
region of $\xi$ the spectrum of the scalar field exhibits at least
one tachyon mode, while for larger values of $\xi$ it is possible to
have two or more tachyon modes, see Appendix A} for $\xi>\xi_{c}$.
The existence of tachyon modes for $\xi<0$ or $\xi>\xi_{c}$ implies
an instability for both the RS2-metric and the scalar field
($\phi=0$) (see also \cite{FP}). The question that arises is whether
there exist other new static stable solutions where the system can
relax. As we discussed in the previous paragraph this model indeed
possesses static solutions different from the standard one of the
RS2-model. The stability and possible implications of these
numerical solutions are discussed in the main part of this paper.

The spectrum of a nonminimally coupled bulk scalar field in the case
of RS1-model \cite{Ran} has been also considered \cite{Toms,Dav},
where it was found that the tachyon character of the model remains
in appropriate regions of $\xi$. However, the authors of Refs.
\cite{Toms,Dav} are not interested to examine the stability of the
RS-model, or to solve the Einstein equations. They mainly use this
result in order to put the standard model on the brane (negative
tension brane) with a bulk Higgs field and a gravity-induced Higgs
mechanism.

In the case of four dimensions the same nonminimally coupled model
has been considered in Refs. \cite{Haw,ford} in the background of de
Sitter space-time. It is obtained that for a specific range of
values of $\xi$ the scalar field is rendered unstable, and this
result has straightforward implications to the cosmological constant
problem.

\section{RS2-model with a nonminimally coupled scalar field}

The action which describes the RS2-model, if we set $8\pi G_{5}=1$
($G_5$ is the five-dimensional Newton constant), is
\begin{equation}
S=\int d^5 x{\cal L}_{RS}=\int d^5
x\left(\frac{1}{2}\sqrt{|g|}\left(
R-2\Lambda\right)-\sigma\delta(z)\sqrt{|g^{(brane)}|}\right)
\end{equation}
where $d^{5}x=d^{4}x dz$, and $z$ parameterizes the extra dimension.
In addition $R$ is the five-dimensional Ricci scalar, $g$ is the
determinant of the five-dimensional metric tensor $g_{MN}$ ($M,
N=0,1,...,4$), and $g^{brane}$ is the determinant of the induced
metric on the brane. We adopt the mostly plus sign convention for
the metric \cite{R2}.

If the fine tuning $\Lambda=\frac{-\sigma^2}{6}$ is satisfied, the
Einstein equations have a \textit{stable static solution} of the
form
\begin{equation}
ds^{2}=e^{-2k|z|}(-dx_{0}^{2}+dx_{1}^{2}+dx_{2}^{2}+dx_{3}^{2})+dz^{2}
\end{equation}
where $k=\sqrt{\frac{-\Lambda}{6}}$.

In this work we aim at the studying of the RS2-model with a
nonminimally coupled bulk scalar field. The action of this model is:
\begin{equation}
 S=\int d^{5}x \;({\cal L}_{RSF}+{\cal L}_{\phi})
\end{equation}
The gravity part of the lagrangian is given by the equation
\begin{equation}
{\cal L}_{RSF}=\sqrt{|g|}\left(
F(\phi)R-\Lambda\right)-\sigma\delta(z)\sqrt{|g^{(brane)}|}
\end{equation}
where the factor
\begin{equation}
F(\phi)=\frac{1}{2}(1-\xi \phi^2 )
\end{equation}
corresponds to a nonminimally coupled scalar field with an
interaction term of the form  ${\cal L}_{int}=-\frac{1}{2}\xi R\;
\phi^{2}$, and $\xi$ is a dimensionless coupling.

The scalar field part of the lagrangian is
\begin{eqnarray}
{\cal L}_{\phi}&=&\sqrt{|g|}\left(-\frac{1}{2}g^{M N} \nabla_{M}\phi
\nabla_{N}\phi-V(\phi)\right)
\end{eqnarray}
where the potential is assumed to be of the standard form
$V(\phi)=\lambda \phi^{4}$.

The Einstein equations, which correspond to the action of Eq. (3)
are
\begin{eqnarray}
G_{MN}+\Lambda\; g_{MN}+\sigma
\delta(z)\frac{\sqrt{|g^{(brane)}|}}{\sqrt{|g|}}g_{\mu\nu}\delta_{M}^{\mu}\delta_{N}^{\nu}=T^{(\phi)}_{MN}
\end{eqnarray}
where the energy momentum tensor for the scalar field is
\begin{equation}
T^{(\phi)}_{MN}=\nabla_{M}\phi\nabla_{N}\phi-g_{MN}[\frac{1}{2}g^{P\Sigma}\nabla_{P}\phi\nabla_{\Sigma}\phi+V(\phi)]+2
\nabla_{M}\nabla_{N}F(\phi)-2 g_{MN}\Box F(\phi)+(1-2 F(\phi))G_{MN}
\end{equation}

The equation of motion for the scalar field is
\begin{equation}
\Box \phi+\frac{\partial F(\phi)}{\partial \phi} R-\frac{\partial
V(\phi)}{\partial \phi}=0
\end{equation}
The above equation is not independent of the Einstein equations (7),
as it is equivalent to the conservation equation $\nabla^M
T^{(\phi)}_{MN}=0$, where $T^{(\phi)}_{MN}$ is given by Eq. (8).

We are looking for static solutions of the form
\begin{equation}
ds^{2}=a^{2}(z)(-dx_{0}^{2}+dx_{1}^{2}+dx_{2}^{2}+dx_{3}^{2})+dz^{2},
\quad \phi=\phi(z)
\end{equation}
From Einstein Equations (Eq. (7)) we obtain two independent
equations:
\begin{eqnarray}
&&G_{ii}+\Lambda\; g_{ii}+\sigma \delta(z)g_{ii}=T^{(\phi)}_{ii}\\
&&G_{zz}+\Lambda\; g_{zz}=T^{(\phi)}_{zz}
\end{eqnarray}
where $i=0,1,2,3$. If we set $a(z)=e^{A(z)}$ and use Eqs.
(8),(10),(11) and (12) we get
\begin{eqnarray}
F_1=0:&&3(1-\xi\phi^{2}(z))\left(A''(z)+2
A'(z)^2\right)+\Lambda+(\frac{1}{2}-2
\xi)\phi'(z)^2+V(\phi(z))\nonumber\\&&-2 \xi \phi(z) \phi''(z)-6 \xi
A'(z)\phi(z) \phi'(z)+\sigma \delta(z)=0
\end{eqnarray}
\begin{eqnarray}
F_2=0:&&6(1-\xi\phi^{2}(z))
A'(z)^2+\Lambda-\frac{1}{2}\phi'(z)^2+V(\phi(z))-8 \xi A'(z)\phi(z)
\phi'(z)=0
\end{eqnarray}
From Eq. (9) for the scalar field we get
\begin{equation}
F_3=0:\quad-\phi''(z)-4 A'(z) \phi'(z)-\xi\left(8
A''(z)+20A'(z)^2\right)\phi(z)+V'(\Phi)=0
\end{equation}
We can show that Eq. (15) (or equation $F_3=0$) can be found from
Eqs. (13) and (14) (or equations $F_1=0$ and $F_2=0$) by taking the
combination $-4A'(z)(F_1-F_2)+F_2'=0$. Note that the relation
$-4A'(z)(F_1-F_2)+F_2'=0$ is equivalent with the condition $\nabla^M
T^{(\phi)}_{MN}=0$.

As we have already mentioned, Eqs. (13),(14) and (15) are not
independent. The solutions can be obtained by integrating the second
order differential equations (13) and (15). The first order
differential equation (14) is an integral of motion of Eq. (13) and
(15), and acts as a constraint between $A(z)$, $\phi(z)$ and their
first derivatives.

In particular we choose to solve the second order differential
equation $F_1-F_2=0$ (see Eqs. (13) and (14)) and $F_3=0$ (see Eq.
(15)), or
\begin{eqnarray}
3(1-\xi\phi^{2}(z)) A''(z)+(1-2 \xi)\phi'(z)^2-2 \xi \phi(z)
\phi''(z)+2 \xi A'(z)\phi(z) \phi'(z)+\sigma \delta(z)=0
\end{eqnarray}
\begin{equation}
-\phi''(z)-4 A'(z) \phi'(z)-\xi\left(8
A''(z)+20A'(z)^2\right)\phi(z)+V'(\Phi)=0
\end{equation}
As this system is complicated we will not look for analytical
solutions, but we will try to solve it numerically. For the
numerical integration of Eqs. (16) and (17) it is necessary to know
the values of $A(0)$, $\phi(0)$, $A'(0^+)$ and $\phi'(0^+)$. These
values are determined by the junction conditions (see Eqs. (18) and
(19) below) and the constraint of Eq. (14).

The delta function in Eq. (16) implies that the first derivatives of
$A(z)$ and $\phi(z)$ are discontinuous on the brane (z=0). If we
integrate Eqs. (16) and (17) over z, in an infinitesimal  interval
$[-\epsilon,\epsilon]$, we get the junction conditions:
\begin{eqnarray}
&&3(1-\xi\phi^{2}(0)) \left(A'(0^+)-A'(0^-)\right)-2\xi
\phi(0)\left(\phi'(0^+)-\phi'(0^-)\right)+\sigma=0\\
&&
-\left(\phi'(0^+)-\phi'(0^-)\right)-8\xi\left(A'(0^+)-A'(0^-)\right)
\phi(0)=0
\end{eqnarray}
If we take into account the $Z_{2}$ symmetry we found
$A'(0^+)-A'(0^-)=2A'(0^+)$ and $\phi'(0^+)-\phi'(0^-)=2\phi'(0^+)$,
thus
\begin{eqnarray}
&&6(1-\xi\phi^{2}(0)) A'(0^+)-4\xi
\phi(0)\phi'(0^+)+\sigma=0\\
&&\phi'(0^+)+8\xi A'(0^+) \phi(0)=0
\end{eqnarray}
By solving these equations with respect to $A'(0^+)$ and to
$\phi'(0^+)$ we obtain
\begin{equation}
A'(0^+)=\frac{-\sigma}{(6-6 \xi \phi(0)^2+32 \xi^2 \phi(0)^2)}
\end{equation}
\begin{equation}
\phi'(0^+)=\frac{8 \xi \sigma \phi(0)}{(6-6 \xi \phi(0)^2+32 \xi^2
\phi(0)^2)}
\end{equation}
However, the problem has an additional constraint for the first
derivatives of $A(z)$ and $\phi(z)$, which is given by the first
order differential equation (14) (or $F_2=0$). Thus if we replace
Eqs. (22),(23) in Eq. (14) we obtain the following sixth order
algebraic equation for $\phi(0)$:
\begin{equation}
\frac{\sigma^2}{(6-6 \xi \phi(0)^2+32 \xi^2
\phi(0)^2)}+\Lambda+V(\phi(0))=0
\end{equation}
Note that if we set $\phi(0)^2=x$ in Eq. (24) we obtain a third
order algebraic equation (if $V(\phi)=\lambda \phi^4$). If
additionally the fine tuning $\Lambda=\frac{-\sigma^2}{6}$ is
satisfied, the constant term of the third order algebraic equation
(24) is zero, and thus we have a second order algebraic equation to
solve, in order to determine the value of $\phi(z)$ on the brane.

\section{Numerical results}

\begin{figure}[h]
\begin{center}
\includegraphics[scale=1,angle=0]{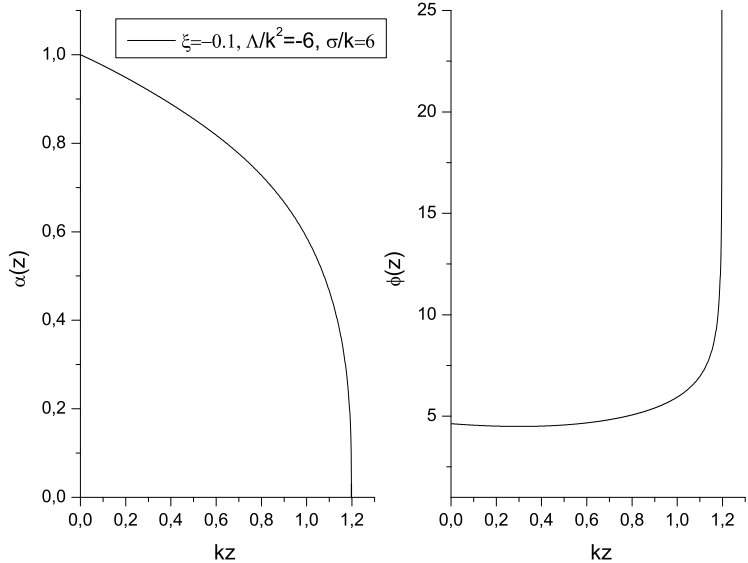}
\end{center}
\caption {The warp factor $a(z)$ and the scalar field $\phi(z)$, in
case (a) ($\xi<0$), as a function of kz for $\sigma/k=6$,
$\Lambda/k^2=-6$, $\xi=-0.1$, $\lambda/k^2=0.01$. The values for
$\sigma,\Lambda$ satisfy the fine tuning of the RS-model. We see
that the metric of our model exhibits a naked singularity, as the
warp factor vanishes for $k z_{s}=1.197$.} \label{1}
\end{figure}

\begin{figure}[h]
\begin{center}
\includegraphics[scale=1,angle=0]{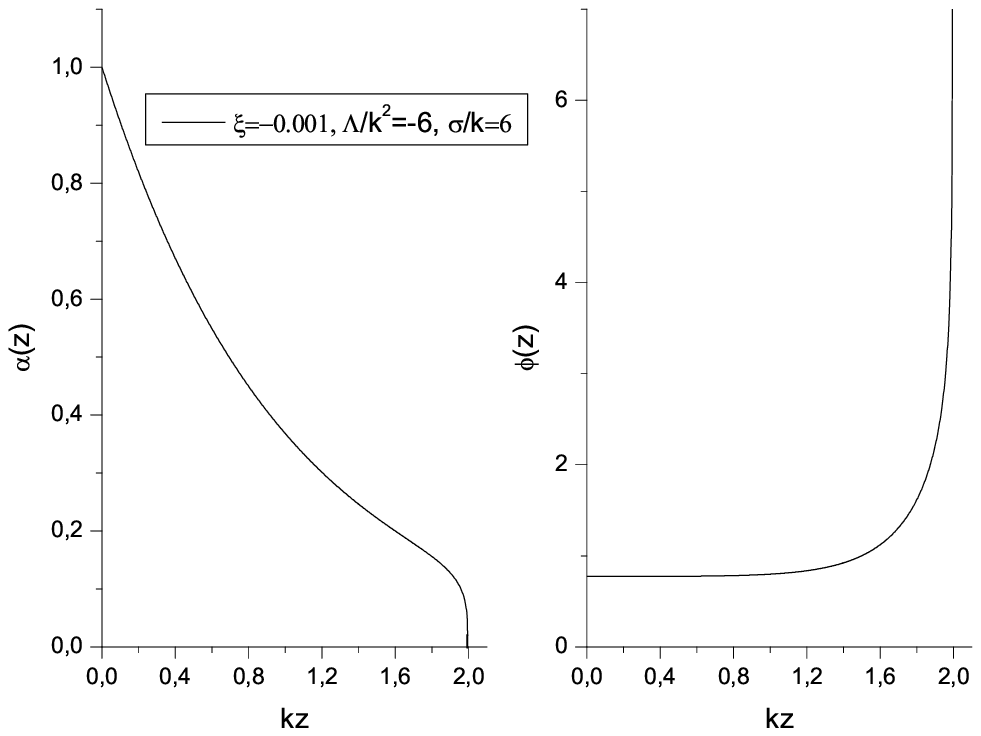}
\end{center}
\caption {The warp factor $a(z)$ and the scalar field $\phi(z)$, in
case (a) ($\xi<0$), as a function of kz for $\sigma/k=6$,
$\Lambda/k^2=-6$, $\xi=-0.001$, $\lambda/k^2=0.01$. The values for
$\sigma,\Lambda$ satisfy the fine tuning of the RS-model. We see
that the metric of our model exhibits a naked singularity, as the
warp factor vanishes for $k z_{s}=1.994$.} \label{2}
\end{figure}

\begin{figure}[h]
\begin{center}
\includegraphics[scale=1,angle=0]{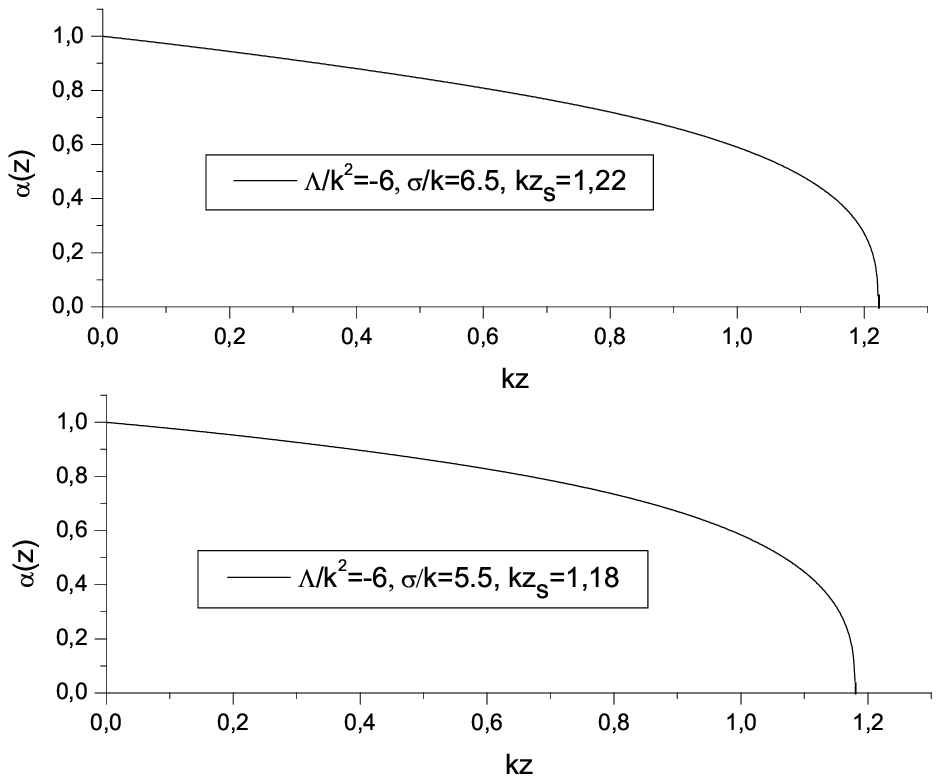}
\end{center}
\caption {The warp factor $a(z)$, in case (a) ($\xi<0$),  as a
function of kz for two values of $\sigma/k=6.5$ and $5.5$,
$\Lambda/k^2=-6$, $\xi=-0.1$, $\lambda/k^2=0.01$. The values for
$\sigma,\Lambda$ do not satisfy the fine tuning of the RS-model.
Note that even when the fine tuning $\Lambda=-\frac{\sigma^{2}}{6}$
is violated the induced metric on the brane is Minkowski.} \label{3}
\end{figure}

For the determination of the two unknown functions $A(z)$ and
$\phi(z)$ we can solve the system of second order differential
equations (16) and (17) for $z\geq 0$ numerically. In order to
integrate it is necessary to know the values of $A(0)$, $\phi(0)$,
$A'(0^+)$ and $\phi'(0^+)$. If we assume that the warp factor is
normalized to unity on the brane (or $a(0)=1$) we find that $A(0)=0$
(note that $a(z)=e^{A(z)}$). The value of $\phi(0)$ is obtained by
solving Eq. (24), and the values of $A'(0^+)$ and $\phi'(0^+)$ can
be found from Eqs. (22) and (23). Then it is an easy task to use a
routine of Fortran or Mathematica to extract the numerical
solutions. Note that the numerical results we obtain satisfy with a
very good accuracy also the first order differential equation (14).

The model we examine has four independent parameters
$\xi,\lambda,\Lambda,\sigma$. We will keep fixed the parameters
$\lambda,\Lambda,\sigma$, assuming that the fine tuning
$\Lambda=-\frac{\sigma^2}{6}$ is satisfied, and we will vary the
parameter $\xi$. As we discuss in the following sections, depending
on the value of $\xi$ we find three classes of numerical solutions
with different characteristics. Also we investigate what happens
when the fine tuning is violated, and we find that in appropriate
regions of the parameter space, the three classes of solutions we
described are preserved. However, there are regions of the
parameters where there are no static solutions (or Eq. (24) has no
real solutions). A thorough investigation of these regions is very
extended and it is beyond the scope of this paper.

It is convenient in numerical analysis to perform the rescaling
$y=kz$ where $k=\sqrt{-\Lambda/6}$. Then the Einstein equations
(13), (14) remain unchanged if the parameters $\Lambda$ and $\sigma$
are divided by $k^2$ and $k$ correspondingly. Note that
$\Lambda/k^2=-6$. The equation (15) for the scalar field remains
unchanged if the parameter $\lambda$ is divided by $k^2$.

We emphasize that the system of Eqs. (16) and (17) with the boundary
conditions of Eq. (22),(23) and (24), in the case of the fine-tuning
$\Lambda=-\frac{\sigma^2}{6}$, has an obvious analytic solution of
the form of Eq. (10) with an exponential warp factor
$a(z)=e^{-k|z|}$ and scalar field vacuum equal to zero $\phi(z)=0$.
This solution is identical to the well known solution of Eq. (2) for
the RS2-model. However, it is unstable for $\xi<0$ and $\xi>\xi_{c}$
(see Ref. \cite{FP} and Appendix A), thus it is worth investigating
whether this model posses static solutions other than that of Eq.
(2), as we do in the rest of this section.

\subsection{Case (a) ($\xi<0$)}

The main feature of the numerical solutions for $\xi<0$ is a naked
singularity at finite proper distance $z_{s}$ in the bulk. The
scalar field $\phi(z)$ is almost constant near the brane, and tends
to infinity as z tends to the singularity point in the bulk.

In Fig. \ref{1} we have plotted the warp factor $a(z)$ and the
scalar field $\phi(z)$ as a function of z for $\sigma/k=6$,
$\Lambda/k^2=-6$, $\xi=-0.1$, $\lambda/k^2=0.01$. The values for
$\sigma,\Lambda$ satisfy the fine tuning of the RS-model. We see
that the metric of our model exhibits a naked singularity, as the
warp factor vanishes for $k z_{s}=1.197$. Note that the warp factor
for $\xi=-0.1$ is completely different from the exponential profile
of the warp factor of the RS2-model.

However for very small absolute values of $\xi$ (for example
$\xi=-0.001$), as we see in Fig. \ref{2}, the warp factor near the
brane is almost identical with the exponential profile of Eq. (2)
($a(z)=e^{-kz}$), (in Fig. \ref{2} we we have plotted the warp
factor $a(z)$ and the scalar field $\phi(z)$ as a function of z for
$\sigma/k=6$, $\Lambda/k^2=-6$, $\xi=-0.001$, $\lambda/k^2=0.01$ and
$k=1$). However even in this case we can not avoid a naked
singularity in the bulk at $z_{s}k=1.994$. We have checked that a
naked singularity in the bulk appears for arbitrarily small negative
values of $\xi$.

In Fig. \ref{3} we examine the case where we have a violation of the
fine tuning $\Lambda \neq-\frac{\sigma^2}{6}$. As a consequence, the
RS2-metric of Eq. (2) (exponential warp factor) is not a solution of
the RS2-model with a nonminimally coupled bulk scalar field.
However, we see (Fig. \ref{3}) that this model possesses static
solutions of the form of Eq. (10) with a naked singularity in the
bulk. It is interesting to note that the induced metric on the brane
is Minkowski, even when the fine tuning
$\Lambda=-\frac{\sigma^{2}}{6}$ is violated. We remind the reader
that in the case of a violation of the fine tuning, the RS2-model
(without the nonminimally coupled scalar field) possesses solutions
with an $AdS_4$ or $dS_4$ induced metric on the brane, see for
example Ref. \cite{AP}. Of course if we wish to construct a brane
world scenario, we should put a second brane in the bulk before the
singularity. However, in this case the tension of the second brane
$\sigma'$, and the parameters $\Lambda$ and $\sigma$ of the first
brane should be finely tuned suitably, if we want the Einstein
equations and the boundary conditions on the second brane to be
satisfied (see for example Ref. \cite{Rub} and references therein).

Note that the numerical solutions we found for $\xi<0$ are similar
to the analytic solutions in Ref. \cite{AH}. However, the model in
Ref. \cite{AH} is quite different from that we assume here. In Ref.
\cite{AH} it is argued that this kind of solutions may have physical
interest even without a second brane. In particular, there is a
possibility that quantum gravity effects near the singularity push
the singularity to infinite proper distance in the bulk. In this
way, the singularity acts effectively as a physical end to the extra
dimension z (for more details see Ref. \cite{AH}). A way to take
into account quantum gravity effects, is to add to the gravity
action of Eq. (1) a Gauss-Bonnet term. We have performed numerical
computations also for this case and the answer is that Gauss-Bonnet
gravity can not resolve the naked singularity (see also \cite{NM}).
The naked singularity remains for arbitrary values (positive or
negative) of the free parameter in front of the Gauss-Bonnet term.

In our previous work \cite{FP} we pointed out the instability of
both the RS2-metric of Eq. (2) and the scalar field vacuum $\phi=0$,
for $\xi<0$. Responsible for this instability is the existence of a
unique tachyon mode localized on the brane. By using this result in
Ref \cite{FP} we tried to guess the form of the new static stable
solution where the system is expected to relax. In particular, we
supposed that the profile of the scalar field vacuum in the bulk is
proportional to the wave function of the tachyon mode (or the scalar
field vacuum is nonzero on the brane and tends rapidly to zero in
the bulk). If we take for granted that $\phi(z)\rightarrow 0$ for
$|z|\rightarrow +\infty$, the warp factor for $|z|\rightarrow
+\infty$ must be of the form $a(z)=e^{-k|z|}$ or $a(z)=e^{k|z|}$, as
only these two functions satisfy the Einstein equations if we set
$\phi(z)=0$. In Ref. \cite{FP} we guess the most plausible behavior
for the new static stable metric. However in this work, by solving
numerically the Einstein equations with the appropriate boundary
conditions, we obtain a completely different behavior for the warp
factor and the scalar field vacuum. The scalar field vacuum has not
the profile of the tachyon mode. As we see in Fig. \ref{1} and Fig.
\ref{2}, the warp factor vanishes in finite proper distance in the
bulk, creating a naked singularity, whereas the scalar field vacuum
tends to infinity as we approach the naked singularity.

Another way to try to resolve the naked singularity is to include a
suitable energy momentum tensor $T_{MN}$ in the right hand side of
Einstein equations (7). We have observed that an energy momentum
tensor of the form $T_{zz}=c/a(z)^4$, with all the other components
equal to zero, is possible to give stable solutions of the form we
described in the previous paragraph. However, we could not find a
physical explanation for an energy momentum tensor of this form.

\subsection{Case (b) ($\xi>\xi_{c}$)}

\begin{figure}[h]
\begin{center}
\includegraphics[scale=1,angle=0]{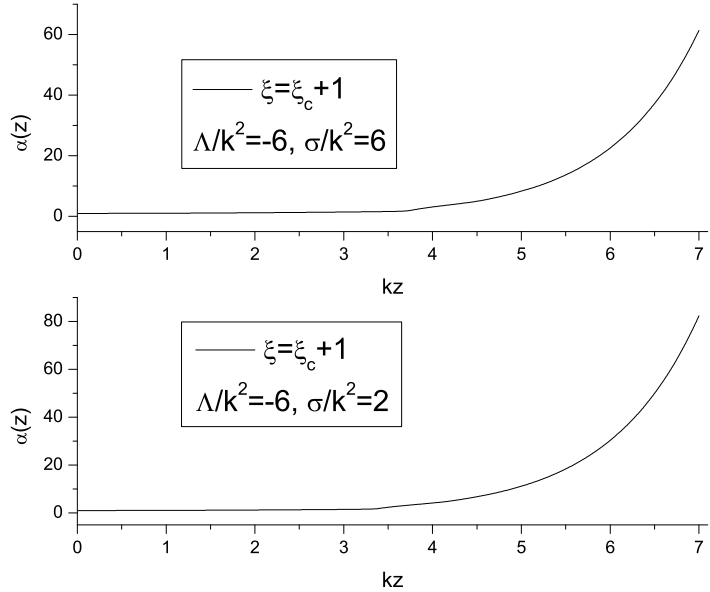}
\end{center}
\caption {The warp factor $a(z)$, in case (b) ($\xi>\xi_c$),  as a
function of z for two values of $\sigma/k=6$ and $2$,
$\Lambda/k^2=-6$, $\xi=\xi_c+1$, $\lambda/k^2=0.01$. Note that the
solution exists even when the values of $\sigma$ and $\Lambda$ do
not satisfy the fine tuning of the RS-model.} \label{4}
\end{figure}

\begin{figure}[h]
\begin{center}
\includegraphics[scale=1,angle=0]{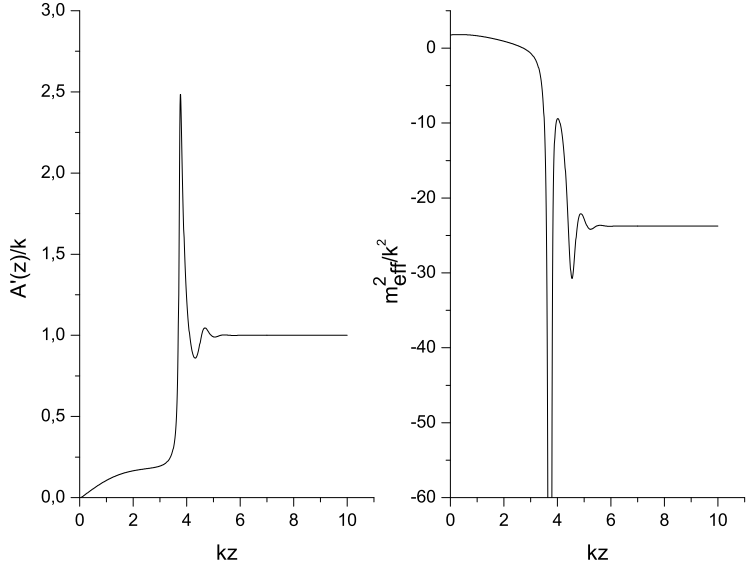}
\end{center}
\caption {The $A'(z)/k$ and $m_{eff}^2/k^2$, in case (b)
($\xi>\xi_c$), as a function of kz for $\sigma/k=6$,
$\Lambda/k^2=-6$, $\xi=\xi_c+1$, $\lambda/k^2=0.01$. Note that
asymptotically both the derivative $A'(z)$ and the $m_{eff}^2$ tend
to a constant value, or the space-time is asymptotically $AdS_5$.}
\label{7}
\end{figure}

\begin{figure}[h]
\begin{center}
\includegraphics[scale=1,angle=0]{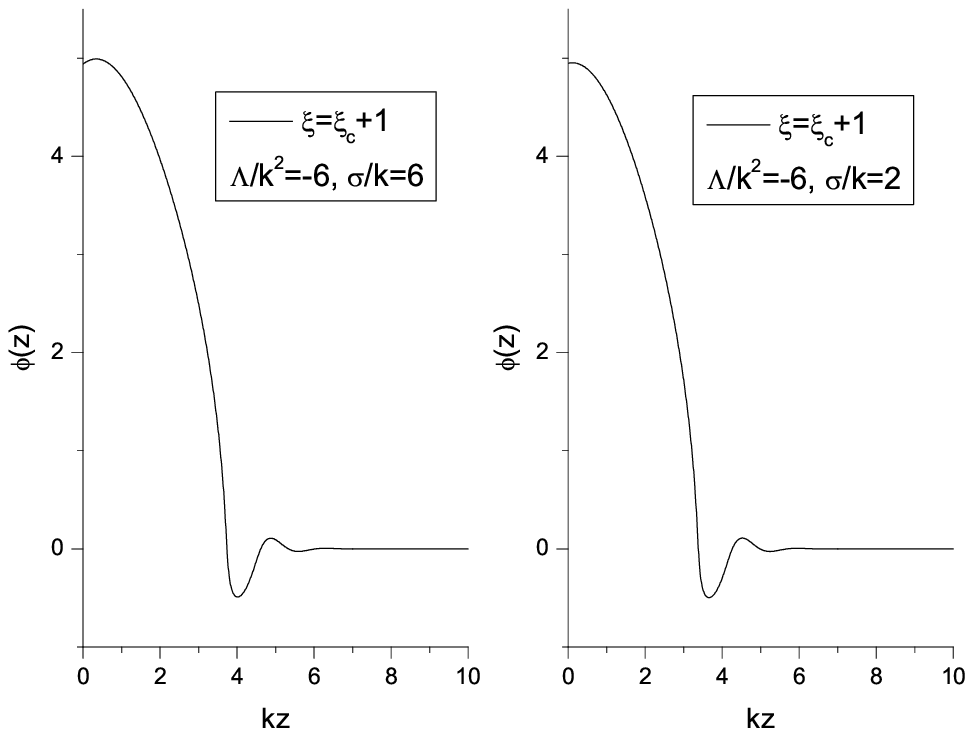}
\end{center}
\caption {The scalar field $\phi(z)$, in case (b) ($\xi>\xi_c$), as
a function of kz for two values of $\sigma/k=6$ and $2$,
$\Lambda/k^2=-6$, $\xi=\xi_c+1$, $\lambda/k^2=0.01$. Note that the
solution exists even when the values of $\sigma$ and $\Lambda$ do
not satisfy the fine tuning of the RS-model.} \label{5}
\end{figure}

For $\xi>\xi_{c}$ we obtain a different class of numerical
solutions. As we see in Fig. \ref{4}, the warp factor $a(z)$ is of
the order of unity in a small region near the brane and increases
exponentially ($a(z)\rightarrow e^{k z}$), as $z\rightarrow
+\infty$. From this figure we can estimate that the warp factor
$a(z)$ is of the order of unity for $0\leq z< 3.5$ and the pure
exponential behavior $a(z)=e^{k z}$ begins for $z> 5$. The
exponential behavior of the warp factor is confirmed with great
accuracy by numerical computations, as in the left-hand panel of
Fig. \ref{7} we observe that $A'(z)\sim 1$ for large values of z.
The scalar field $\phi(z)$, as we see in Fig. \ref{5}, is nonzero on
the brane (in a rather small region ($0\leq kz< 4$), and for $kz> 5$
tends rapidly to zero. We have checked carefully that this class of
solutions appears even when $\xi$ exceeds by a small amount the
conformal coupling $\xi_{c}$ ($\xi=\xi_c+\delta\xi$ with
$\delta\xi<<1$).

Note that this class of numerical solutions is preserved even when
the fine tuning $\Lambda=-\frac{\sigma^2}{6}$ is not satisfied (see
the down panel of Fig. \ref{4} and the right panel of Fig. \ref{5}
($\sigma=2$, $\Lambda=-6$)). In this case also, $A'(z)\sim 1$ for
large values of z. We have checked numerically that for fixed
negative \footnote{Note that for nonnegative values of $\Lambda$
there are no solutions of the form we described.} $\Lambda$ there
are solutions of the form of Figs. \ref{4} and \ref{5} only if the
brane tension $\sigma$ is smaller than the absolute value of the
five dimensional cosmological constant $\Lambda$.

A feature of this class of solutions is that the warp factor tends
to infinity $a(z)\rightarrow e^{k z}$ for large $z$. If we wish to
construct a brane world scenario, it is necessary to include a
second positive tension brane far away from the first, where the
numerical value of the scalar field is practically zero. In this
case the tension $\sigma'$ of the second brane should satisfy the
fine tuning condition $\Lambda=\frac{-\sigma'^2}{6}$.

However the most important topic for the construction of a realistic
brane world scenario is the stability of this class of solutions. In
order to study the stability, we will replace $\phi\rightarrow
\phi+\hat{\phi}$ in the lagrangian
\begin{eqnarray}
{\cal L}&=&\sqrt{|g|}\left(-\frac{1}{2}g^{M N} \nabla_{M}\phi
\nabla_{N}\phi-\frac{1}{2}\xi R \phi^2-V(\phi)\right)
\end{eqnarray}
where $\hat{\phi}$ is a small perturbation around the classic
solution. If we keep only second order terms of $\hat{\phi}$, we
obtain that an effective mass $m_{eff}^{2}=\xi R+12 \lambda
\phi(z)^2$ arises for the scalar field perturbation $\hat{\phi}$. In
the right-hand panel of Fig. \ref{7} we have plotted the effective
mass $m_{eff}^{2}=\xi R+12 \lambda \phi(z)^2 $ as a function of z.
We see that as $z\rightarrow +\infty$ the square of the effective
mass tends to a constant negative value. This result indicates
\footnote{A negative mass term is not enough to establish the
instability of the model. In the case of curved space-time it is
necessary for at least one tachyon mode for the scalar field to
exist. In Appendix B we show that indeed the spectrum of the scalar
field exhibits a tachyon character. In particular, there is a
continuous spectrum of tachyon modes for $\xi>\xi_c$.} an
instability of the solutions of class (b) against scalar field
perturbation. Hence this class of solutions can not be used for the
construction of realistic brane world scenarios.

\subsection{Case (c) ($0<\xi<\xi_{c}$)}

\begin{figure}[h]
\begin{center}
\includegraphics[scale=1,angle=0]{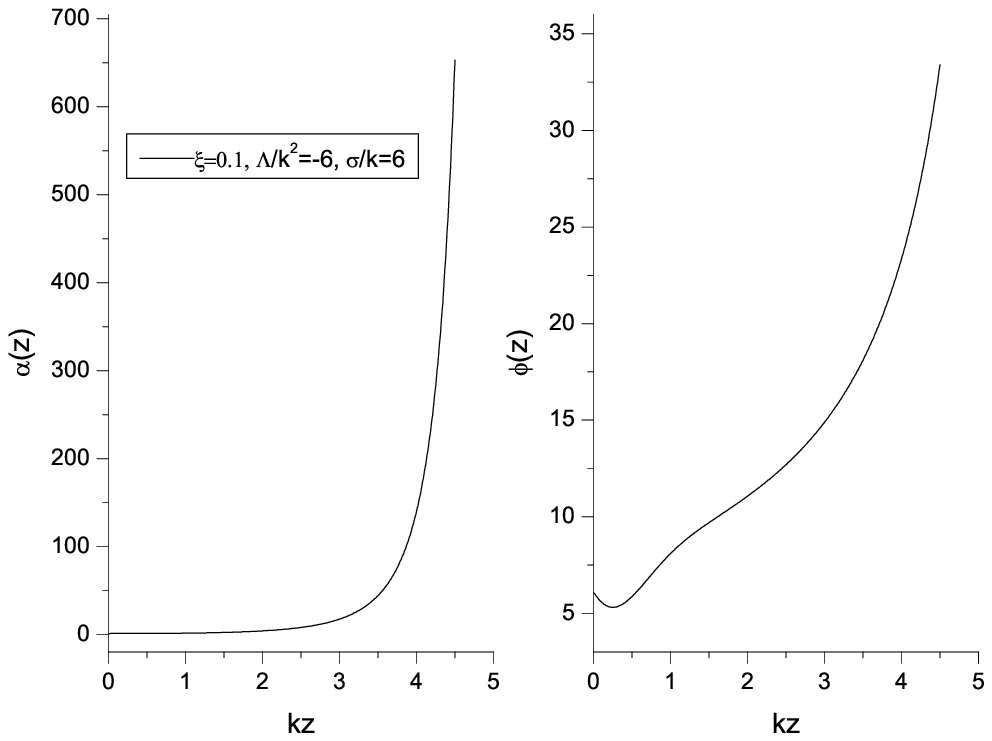}
\end{center}
\caption {The warp factor $a(z)$ and the scalar field $\phi(z)$, in
case (c) ($0<\xi<\xi_c$), as a function of kz for $\sigma=6$,
$\Lambda=-6$, $\xi=0.1$, $\lambda=0.01$. The values for
$\sigma,\Lambda$ satisfy the fine tuning of the RS-model.} \label{6}
\end{figure}

For $0<\xi<\xi_{c}$ the standard RS2-metric is stable, as there are
no tachyon modes for the scalar field perturbations around this
solution (see Appendix A). As we have mentioned, in this case a
second static solution can be found by solving the Einstein
equations numerically. This second solution exists only if Eq. (24)
has at least a nonzero real root for $\phi(0)$. As we see in Fig.
\ref{6} for $0<\xi<\xi_{c}$ the warp factor $a(z)$ of this solution
tends rapidly to infinity (faster than case (b)), also the scalar
field $\phi(z)$ is nonzero on the brane and tends rapidly to
infinity in the bulk. Contrary to case (b), where the space-time is
asymptotically $AdS_{5}$, in this case the scalar curvature tends to
infinity. Solutions of this kind exist even when the fine tuning of
the RS-model is violated.

If we wish to construct a brane world model, we should put a second
brane in the bulk. However, in order to satisfy simultaneously the
three boundary conditions of Eqs. (22),(23) and (24) on the second
brane, it is necessary to assume a fine tuning for the parameters of
the model. Moreover, as we did in case (b), we have computed the
effective mass of the scalar field around this solution and we have
seen that it is positive. This implies that this class of solution
is stable against perturbations of the scalar field.

\section{Minimal coupling and conformal coupling}

In this section we examine separately the special cases of $\xi=0$
and $\xi=\xi_c$. Especially in the case of conformal coupling, when
the fine tuning is violated, we have a physically interesting static
stable solution.

\subsection{Minimal coupling $\xi=0$}

The algebraic equation (24), for $\phi(0)$, can be solved
analytically for $\xi=0$ or $\xi=\xi_c$. If we set $\xi=0$ or
$\xi=\xi_c$ we find the equation $\lambda
\phi(0)^4+\sigma^2/6+\Lambda=0$. In the case of fine tuning we
obtain $\phi(0)=0$, hence the solution of the model we examine is
the standard one with warp factor $a(z)=e^{-k|z|}$ and $\phi(z)=0$.
If $\sigma^2/6+\Lambda>0$, the above equation has no real roots and
our model has no static solutions. On the other hand, for
$\sigma^2/6+\Lambda<0$ we have two real roots
$\phi(0)=\pm\sqrt[4]{-\lambda^{-1}(\sigma^2/6+\Lambda)}$, and in
this case the static numerical solutions for $\xi=0$ have the same
characteristics with those of class (a) ($\xi<0$) with a naked
singularity.

\subsection{Conformal coupling $\xi=\xi_c$}

\begin{figure}[h]
\begin{center}
\includegraphics[scale=1.2,angle=0]{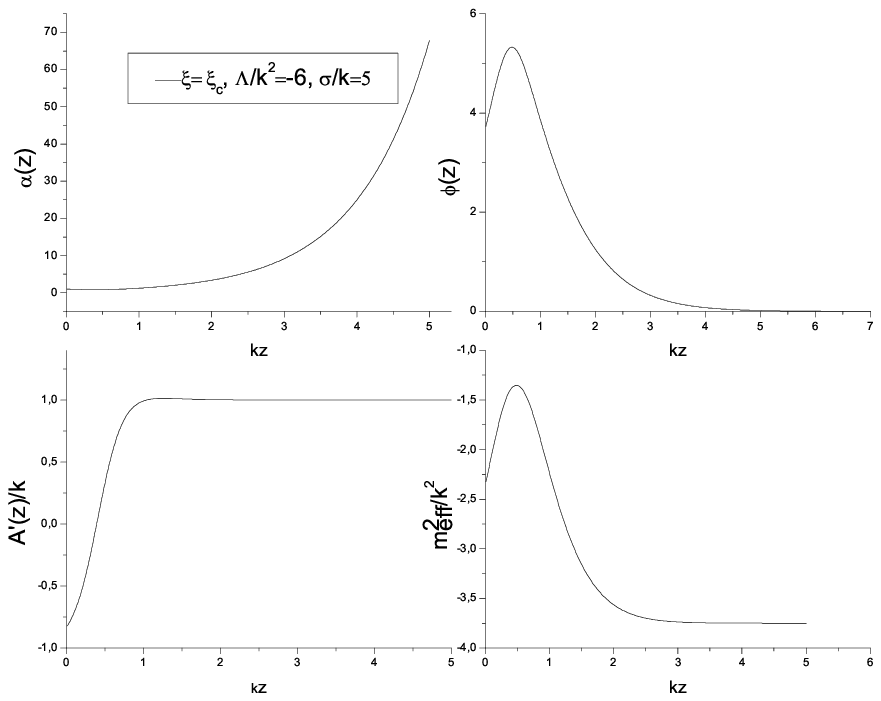}
\end{center}
\caption {$a(z)$, $\phi(z)$, $A'(z)/k$ and $m_{eff}^2(z)/k^2$, for
$\xi=\xi_c$, as a function of kz for $\sigma/k=5$, $\Lambda/k^2=-6$
and $\lambda/k^2=0.01$. Note that asymptotically both the derivative
$A'(z)$ and the $m_{eff}^2$ tend to a constant value, or the
space-time is asymptotically $AdS_5$.} \label{con1}
\end{figure}

\begin{figure}[h]
\begin{center}
\includegraphics[scale=1.2,angle=0]{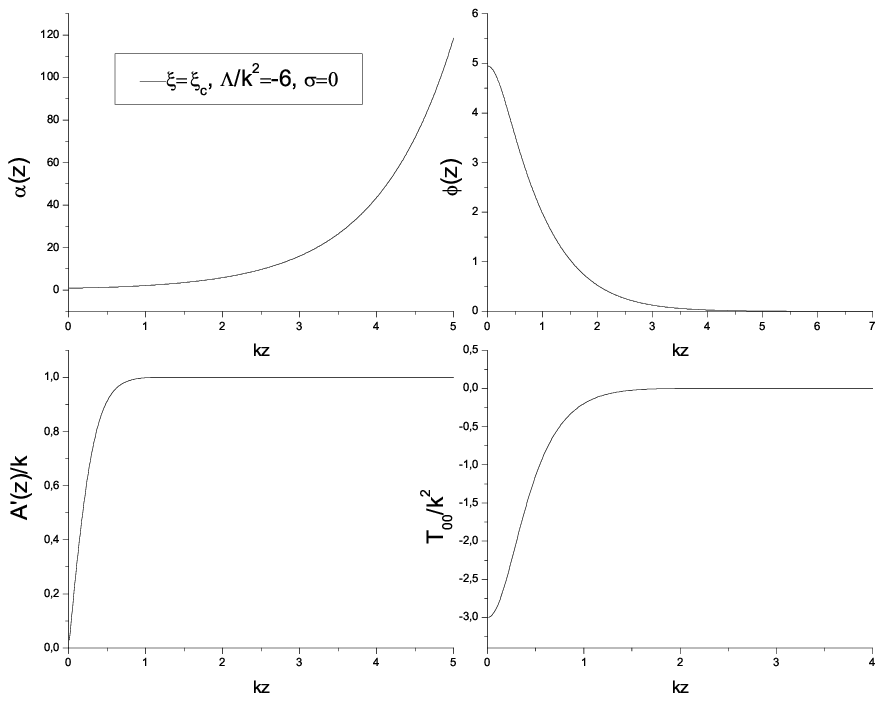}
\end{center}
\caption {$a(z)$, $\phi(z)$, $A'(z)/k$ and $T_{00}(z)/k^2$, for
$\xi=\xi_c$, as a function of kz for $\sigma=0$, $\Lambda/k^2=-6$
and $\lambda/k^2=0.01$.} \label{con2}
\end{figure}

In this Section we examine the model when $\xi=\xi_c$ and
$\Lambda\neq-\frac{\sigma^2}{6}$. As we have already discussed in
Section 4.1, if $\sigma/k<6$, the algebraic equation (24) for
$\phi(0)$ has two real roots $\phi(0)=\pm \phi_0$, where
$\phi_0=\sqrt[4]{-\lambda^{-1}(\sigma^2/6+\Lambda)}$. In this case,
the numerical solutions of the model exhibit the same
characteristics with those of class (b) ($\xi>\xi_c$). In
particular, the warp factor $a(z)$ is of the order of unity near the
brane and increases exponentially ($a(z)\sim e^{k |z|}$ ), as
$z\rightarrow \pm \infty$, while the scalar field $\phi(z)$ is
nonzero on the brane and tends rapidly to zero in the bulk, as we
can see in the upper panels of Fig. \ref{con1}. The exponential
behavior of the warp factor $a(z)=e^{A(z)}$ is confirmed in the
down-left panel of Fig. \ref{con1}. The solutions in the case of
conformal coupling are of the form we described only for suitable
small values of $\lambda/k^2$ depending on the exact value of
$\sigma$ (i.e. for $\sigma=5$ we have $\lambda/k^2\leq 0.1$). For
larger values of $\lambda$ we find a different class of solutions,
however it is not of physical interest.

In the down-right panel of Fig. \ref{con1} we observe that the
effective mass of the scalar field is negative. However the
solutions for $\xi=\xi_c$ are stable, as in Appendix C we find that
the scalar field does not exhibit four dimensional tachyon modes.
The result of stability implies that the solutions  for $\xi=\xi_c$
can be used as the background for the construction of a brane-world
scenario.

Now we will try to discuss possible physical implications of the
solutions in the case of conformal coupling. As we have already
mentioned, the warp factor behaves as $e^{k|z|}$ for large $z$ (see
the down-left panel of Fig. \ref{con1}). It is well known that an
exponentially increasing warp factor comes from negative tension
brane with $-\sigma'/k=-6$. Hence, the positive tension brane at
$z=0$ $(\sigma/k=5)$ together with the negative energy density of
the scalar field vacuum $T^{(\phi)}_{00}$, act effectively as a
negative tension brane, which creates the exponential profile
$e^{k|z|}$ for the warp factor. Note that the energy density of the
scalar field has a finite size extension into the bulk of the order
of $1/k$, as we see in the upper-right panel of Fig. \ref{con1}. The
model is completed if we include a second positive tension brane
with $\sigma'/k=6$, in a position $z_c$ where the scalar field
vacuum is practically zero ($z_c>>5/k$). This two-brane set up is
very similar to the well known RS1-model (see introduction). In
analogy with the RS1-model, we will assume that the standard model
particles live in the effectively negative tension brane, or visible
brane . Note that in our case the visible brane is a formation of
the positive tension brane with $\sigma/k=5$ and the negative energy
density $T^{(\phi)}_{00}$ of the scalar field vacuum.

The advantage of this model is that it incorporates a mechanism for
the localization of standard model particles on the visible brane.
We can enrich our model with a Gauge field symmetry, i.e. SU(5) (see
also \cite{FP}). We assume that this gauge field symmetry (SU(5)) is
spontaneously broken to $SU_c(3)\times SU_L(2)\times U_Y(1)$ near
the brane , due to the nonzero value of the scalar field, while it
is restored in the bulk for $z>>5/k$, where the value of the scalar
field is practically zero. This phase structure, Higgs phase on the
brane and confinement phase in the bulk, triggers the Dvali-Shifman
mechanism \cite{Shif} for localization of Gauge fields. In this way
the gauge fields of $SU_c(3)\times SU_L(2)\times U_Y(1)$ and the
matter fields with gauge charge (see Ref. \cite{Rub}) are localized
on the brane, as for escaping in the bulk, it requires energy equal
to $\Lambda_{gap}$, where $\Lambda_{gap}$ is the mass gap emerging
from the nonperturbative confining dynamics of the SU(5) gauge field
theory in the bulk.

In the case of the RS1-model, when ordinary matter is localized on
the negative tension brane, we have the relation
$M_P^2=\frac{M_{*}^3}{k}(e^{2kz_c}-1)$ (see Ref. \cite{Rub}), where
$M_P=10^{19} GeV$ is the four-dimensional Planck scale, $M_*$ is the
fundamental five-dimensional gravity scale, k is the inverse $AdS_5$
radius, and $z_c$ is the position of the second positive tension
brane in the bulk. We have checked that the above relation is valid
approximately also for the model we examine. If we wish to have
strong gravity at TeV we must choose $M_*=1TeV$. According to this
localization mechanism the particles on the visible brane see an
effective length toward the extra dimension of the order of $5/k$.
If we take into account that the masses of ordinary particles are
smaller than 1TeV, we must choose $k/5>>1TeV$. For example, if
$k=100 TeV$ we obtain $10 M_P=M_* e^{kz_c}$, and thus the position
of the second brane is $z_c\approx 39/k$.

Note that even for a zero tension ($\sigma=0$) brane at $z=0$, we
obtained solutions of the form we described in the previous
paragraphs, see Fig. \ref{con2}. In this case the visible brane is
formed only from the negative energy density of the nonminimally
coupled bulk scalar field, as we see in the down-left panel of Fig.
\ref{con2}. In addition, we have found that the numerical solutions,
for small values of $\lambda/k^2$ ($\lambda/k^2\leq 0.01$), can be
approximated with a very good accuracy from the analytical
expressions $a(z)=\left(\cosh(\frac{
|\Lambda|}{2}k|z|)\right)^\frac{2}{|\Lambda|}$ and
$\phi(z)=\phi(0)/\left(\cosh(\frac{ |\Lambda|}{2}k|
z|)\right)^\frac{1}{2}$. These expressions satisfy the constraint
equation (14). However, their second derivatives are not in
agreement with the second derivatives of the numerical solutions,
hence the equations (16) and (17) are not satisfied by the above
mentioned analytical expressions.

\section{Conclusions and discussion}

In this work we studied for the first time the RS2-model with a
nonminimally coupled bulk scalar field, via an interaction term
${\cal L}_{int}=-\frac{1}{2}\xi R \phi^2$. By solving numerically
the Einstein equations with the appropriate boundary conditions on
the brane, we showed that depending on the value of the nonminimal
coupling $\xi$ this model possesses three classes of new static
solutions with different characteristics.

Class (a) ($\xi<0$) develops a naked singularity in the bulk. Class
(b) ($\xi>\xi_c$) consists of solutions which are characterized by
an exponential warp factor $a(z)\sim e^{k z}$, as $z\rightarrow
+\infty$, and a scalar field $\phi(z)$ which is nonzero on the brane
and tends rapidly to zero in the bulk. The solutions of class (c)
($0<\xi<\xi_c$) are characterized by a very fast increase of the
warp factor (faster than class (b)), and a scalar field $\phi(z)$
which tends rapidly to infinity for large $z$. The cases of minimal
and conformal coupling ($\xi=0$ and $\xi=\xi_c$) have been discussed
separately in Sections 4.1 and 4.2.

An interesting point is that these three classes of solutions exist
even when the standard fine tuning of the RS-model is violated. In
order to construct a brane world model by using these solutions, it
is necessary to assume a second brane in the bulk. However, in this
case we should impose a new fine tuning between the parameters of
the model, if we wish to satisfy the boundary conditions on the
second brane.

In addition, we have examined the stability properties of the new
solutions. The solutions of class (a) ($\xi<0$) and class (b)
($\xi>\xi_c$) are unstable against scalar field perturbations, as in
Appendix B we have found that the spectrum of the scalar field
around these solutions exhibits a tachyon character. The only way to
render these solutions stable is to put a second brane in the bulk
before the potential for the scalar field V(w) becomes negative (for
details see Appendix B). The solutions of class (c) are stable, as
the scalar field spectrum has no tachyon modes.

We emphasize that this work has been motivated by our previous work
of Ref. \cite{FP}, where we had assumed the same model with an
additional gauge field symmetry for the scalar field. In particular,
in that work we have argued for a gravity-induced localization
mechanism for $\xi<0$, which is very useful for the localization of
gauge fields on the brane. In this mechanism we have a specific
phase structure: confinement phase in the bulk and Higgs phase on
the brane. Gauge fields, and more generally fermions and bosons with
gauge charge (see Refs. \cite{Shif,Rub}), can not escape into the
bulk unless we give them energy greater than the mass gap
$\Lambda_{gap}$, which emerges from the nonperturbative confining
dynamics of the gauge field model in the bulk. In Ref. \cite{FP} we
observed that the effective mass $m_{eff}^2=\xi R$, for $\xi<0$, is
negative on the brane and positive in the bulk. This result
indicates a phase structure which can trigger the Dvali-Shifman
mechanism in a gravitational way. We would like to emphasize that
the sign of the effective mass, in the case of curved space-time, is
just an indication for the expected phase structure and is not a
strict proof. A realistic gravity-induced localization mechanism
requires a static stable solution with a nonzero scalar field on the
brane which vanishes rapidly in the bulk. In Ref. \cite{FP} we
assumed the existence of a solution of the form we described above.
However, the correct way to find if a solution of this kind really
exists, is to solve the complete system of Einstein equations, as we
do in this paper. We found that no static stable solution of the
required form exists for $\xi<0$. Additionally, we obtained that for
$\xi<0$ the solutions suffer from a naked singularity in the bulk,
see Figs. \ref{1},\ref{2} and \ref{3}. We tried to resolve this
naked singularity by adding a Gauss-Bonnet term to the gravity
action. We performed numerical computations and we found that the
naked singularity remains even in the case of Gauss-Bonnet gravity.
According to the above analysis, it seems that no realistic
gravity-induced localization mechanism for $\xi<0$ can be
constructed.

Finally, we have examined the case of conformal coupling
($\xi=\xi_c$) when the fine tuning of the RS-model is violated, see
Section 4.1. We obtain that the solutions for $\xi=\xi_c$ have the
same characteristics with those of the solutions of class (b)
($\xi<\xi_c$). However, there is an important difference between
them, as the solutions for $\xi=\xi_c$ are stable against scalar
field perturbation (see Appendix C), contrary to the case of second
class, where the solutions are unstable (see Appendix B ). In
Section 4.2 we argue that this class of static stable solutions can
be used for the construction of a realistic brane world scenario,
which is very similar to the well known RS1-model. In this scenario,
the visible brane is a formation of the positive tension brane
$\sigma$ at $z=0$, plus the negative energy density of the
nonminimally coupled scalar field. The advantage of this model is
that it incorporates a mechanism (gravity-induced Dvali-Shifman
mechanism) for the localization of standard model particles on the
visible brane together with the spontaneous breaking of a Grand
Unified Gauge group. We would like to emphasize that the above
mentioned scenario remains even in the case of a zero tension brane
at $z=0$ ($\sigma=0$). In this case, the visible brane is formatted
only from the negative energy density of the bulk scalar field.

\section{Acknowledgements} We are grateful to Professors A. Kehagias
and G. Koutsoumbas  for reading and commenting on the manuscript. We
also thank Professor K. Tamvakis, Professor N. Mavromatos, Dr. P.
Manouselis, Dr. A. Pappazoglou and Dr. N. Prezas, for important
discussions. This work was supported by the "Pythagoras" project of
the Greek Ministry of Education-European community (EPEAEK-EKT,
25/75).

\appendix

\section{Appendix: Spectrum of scalar field perturbations around the
RS2-vacuum}

\begin{figure}[h]
\begin{center}
\includegraphics[scale=1,angle=0]{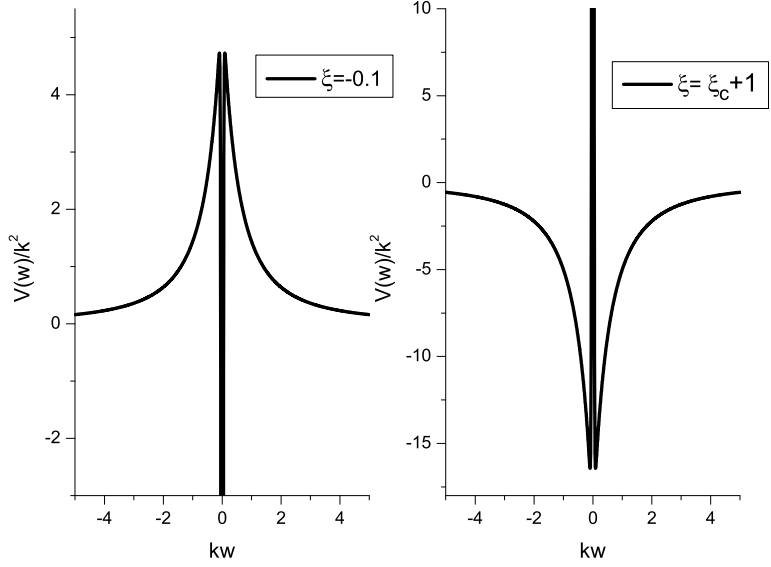}
\end{center}
\caption {The potential $V(w)/k^2$ as a function of $kw$, for
$\xi=-0.1$ (left-hand panel) and $\xi=\xi_c+1$ (right-hand panel).}
\label{8}
\end{figure}

\begin{figure}[h]
\begin{center}
\includegraphics[scale=1,angle=0]{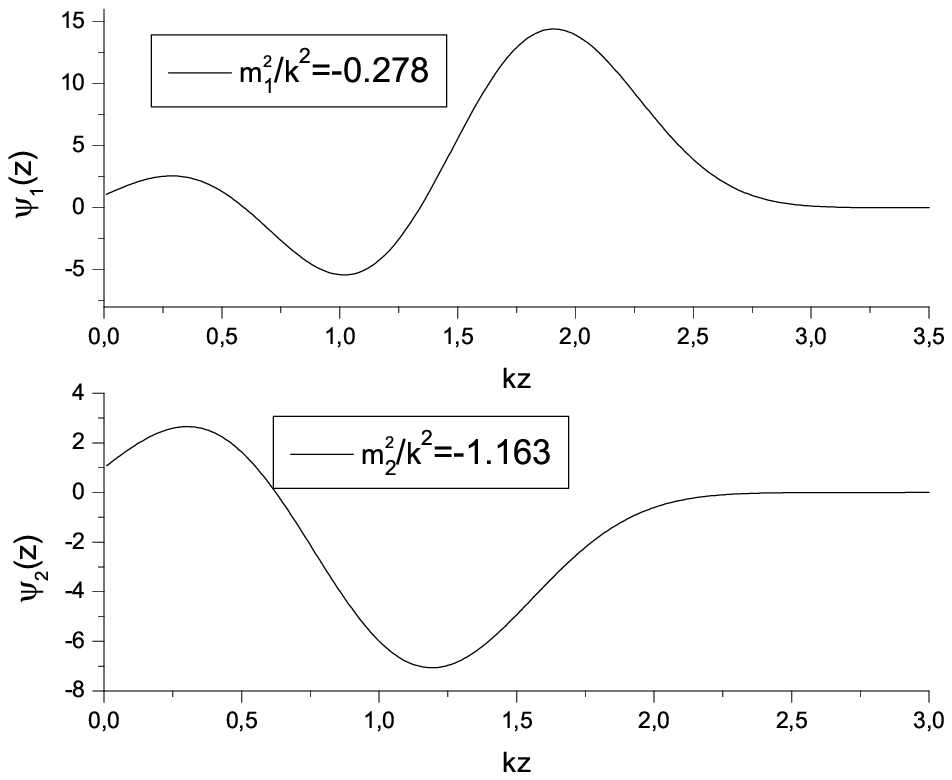}
\end{center}
\caption {For $\xi=\xi_c+1$ the spectrum of the scalar field
contains exactly two tachyon modes with negative energies
$m_{1}^{2}/k^2=-0.278$ and $m_{2}^{2}/k^2=-1.163$. In the figure we
have plotted the wavefunctions $\psi_1$ and $\psi_2$ of the two
tachyon modes as a function of $z$.} \label{9}
\end{figure}

In Ref. \cite{FP} we studied the spectrum of scalar field
perturbations in the background of the RS2-vacuum for $\xi<0$. In
this appendix we complete this investigation by including positive
values for $\xi$.

If we use the transformation
\begin{equation}
w=sgn(z)\frac{(e^{k|z|}-1)}{k}
\end{equation}
the RS2-metric of Eq. (2) can be put into the manifestly conformal
to the five-dimensional Minkowski space form
\begin{equation} ds^{2}=\alpha(w)^2(-dx_{0}^{2}+dx_{1}^{2}+dx_{2}^{2}+dx_{3}^{2}+dw^{2})
\end{equation}
where
\begin{equation} \alpha(w)=
\frac{1}{k|w|+1}
\end{equation}

 The lagrangian of the scalar field if we include an
interaction term $-\frac{1}{2} R \phi^2$ is written as
\begin{equation}
L=\sqrt{|g|}\left(-\frac{1}{2}g^{MN} \nabla_{M}\phi
\nabla_{N}\phi-\frac{1}{2}\xi R\ \phi^2-V(\phi)\right)
\end{equation}
If we consider a small perturbation $\hat{\phi}$ around the scalar
field vacuum ($\phi=0$) we find the corresponding linearized
equation:
\begin{equation}
\frac{1}{\sqrt{|g|}}\;
\partial_{M}\left[\sqrt{|g|}g^{MN}\partial_{N}\hat{\phi}(x,w)\right]+\xi
R(w) \;\hat{\phi}(x,w)=0
\end{equation}

We can set
\begin{equation}
\hat{\phi}(x,w)=e^{ipx} \frac{\psi(w)}{\alpha^{3/2}(w)}
\end{equation}
where $\alpha(w)=1/(k|w|+1)$, and $m^{2}=p_{\mu}p^{\mu}$ is the
effective four dimensional mass.

The function $\psi(w)$ satisfies the Schrondiger like equation
\begin{equation}
-\psi''(w)+\left[V(w)-m^{2}\right]\psi(w)=0
\end{equation}
where the potential $V(w)$ is equal to
\begin{equation}
V(w)=\frac{(\alpha^{3/2}(w))''}{\alpha^{3/2}(w)}+\xi\alpha^{2}(w)R(w)
\end{equation}
From Eqs. (27) and (32) we get
\begin{equation}
V(w)= -16 k(\xi-\xi_c)\left(-\delta(w)+\frac{5
k}{4(k|w|+1)^{2}}\right)
\end{equation}
where $\xi_c=3/16$ is the five dimensional conformal coupling.

Note that the coefficient in front of the potential change sign when
$\xi$ crosses the five dimensional conformal coupling. This result
implies that the potential has two characteristic forms, as we see
in the left-hand panel ($\xi<\xi_c$) and the right-hand panel
($\xi>\xi_c$) of Fig. \ref{8}.

In the first case, where $\xi<\xi_c$ the coefficient of the delta
function is negative. In Ref. \cite{FP} we have shown that if
$\xi<0$, the spectrum of the scalar field contains a unique tachyon
mode localized on the brane (see Eq. (32) in Ref. \cite{FP}). It is
well known that if $\xi=0$ there is no tachyon mode, but there is a
zero mode.

For $0<\xi\leq \xi_c$ the potential still has the form of the
left-hand panel of Fig. \ref{8} (volcano form). However, in this
case the coefficient of the delta function is not large enough in
order to support a tachyon mode. We can prove this result by solving
numerically the eigenvalue equation (Eq. (34) in Ref. \cite{FP}). We
find that there is no tachyon mode in this region of $\xi$.

For $\xi>\xi_c$ the tachyon mode returns, as we obtain from Eq. (34)
in Ref. \cite{FP}. However we can not use Eq. (34) (of Ref.
\cite{FP}) if $\xi>1/5$, as the index of the modified Bessel
functions in Eq. (34)) becomes imaginary. In this case we have
solved numerically the  Schrondiger equation (31) by using a
shooting method and we have confirmed that indeed the spectrum of
the scalar field possesses at least one tachyon mode. In particular
for $\xi=\xi_c+1$ we have exactly two tachyon modes, and in Fig.
\ref{9} we have plotted the corresponding wave functions. Note that
for $\xi>\xi_c$ the potential has the double-well form of the
right-hand panel of Fig. \ref{9}. In this region of $\xi$ the
spectrum of the scalar field is possible to contain one or more
tachyon modes according to the depth of the double well.

In this appendix we studied the spectrum of scalar field
perturbations around the RS2-vacuum. Depending on the value of $\xi$
we obtain that: (a) for $\xi<0$ we have unique tachyon mode, (b) for
$\xi>\xi_c$ we have at least one tachyon mode and (c) for
$0<\xi<\xi_c$ there are no tachyon modes. We conclude that the
RS2-vacuum is unstable against scalar field perturbations in cases
(a) and (b). For $\xi=0$ or $\xi=\xi_c$ there are no tachyon modes,
hence in these cases the RS-2 vacuum is stable.

\section{Appendix: The tachyon character of the spectrum around the
solutions of first and second class}

\begin{figure}[h]
\begin{center}
\includegraphics[scale=1,angle=0]{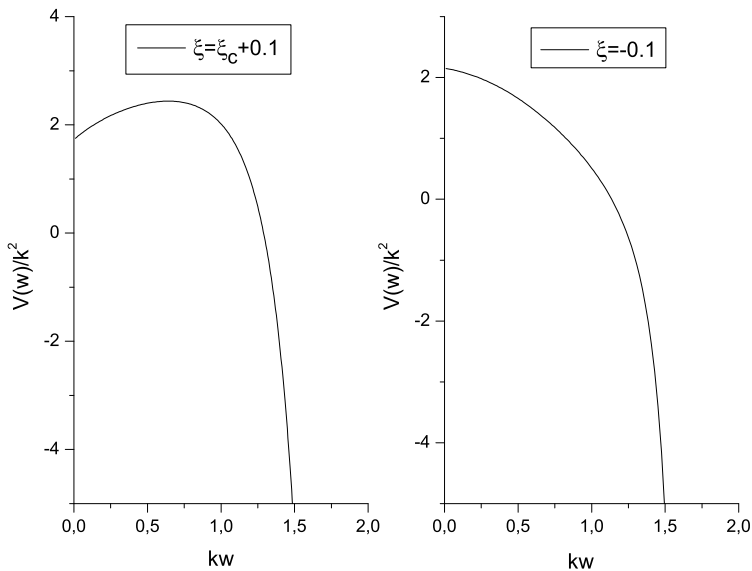}
\end{center}
\caption {The potential $V(w)/k^2$ as a function of $kw$, for
$\xi=\xi_c+0.1$ in the background of second class solutions
(left-hand panel) and for $\xi=-0.1$ in the background of first
class solutions (right-hand panel), for $\sigma/k=6$,
$\Lambda/k^2=-6$, ,$\lambda/k^2=0.01$.} \label{10}
\end{figure}

In this appendix we will show that the spectrum of scalar field
perturbation around the solutions of first class ($\xi<0$) and
second class ($\xi>\xi_c$) exhibits a tachyon character. We can
follow the analysis that is presented in the previous appendix. The
only difference is that the value of the scalar field is nonzero in
the bulk, hence the potential that appears in the Schrondiger
equation (32) must be modified as
\begin{equation}
V(w)=\frac{(\alpha^{3/2}(w))''}{\alpha^{3/2}(w)}+\alpha^{2}(w)\left(\xi
R(w)+12\lambda \phi^2(w)\right)
\end{equation}
where $w=\int_{0}^{z}dz/a(z)$.

In right-hand panel and left-hand panel of Fig. \ref{10} we have
plotted the potential $V(w)$ for the first class of solutions
$\xi<0$, and the second class of solutions $\xi>\xi_c$ respectively.
As we see, the potential for both cases becomes negative and
decreases without a lower bound. This implies a continuous spectrum
of tachyon modes, and as a result both the first and second class of
solutions are unstable. A possible way to resolve this instability
is to put a second brane before the potential $V(w)$ becomes
negative. In this case, a fine tuning between the parameters of the
model is necessary.

\section{Appendix: Stable solutions for $\xi=\xi_c$}

\begin{figure}[h]
\begin{center}
\includegraphics[scale=1,angle=0]{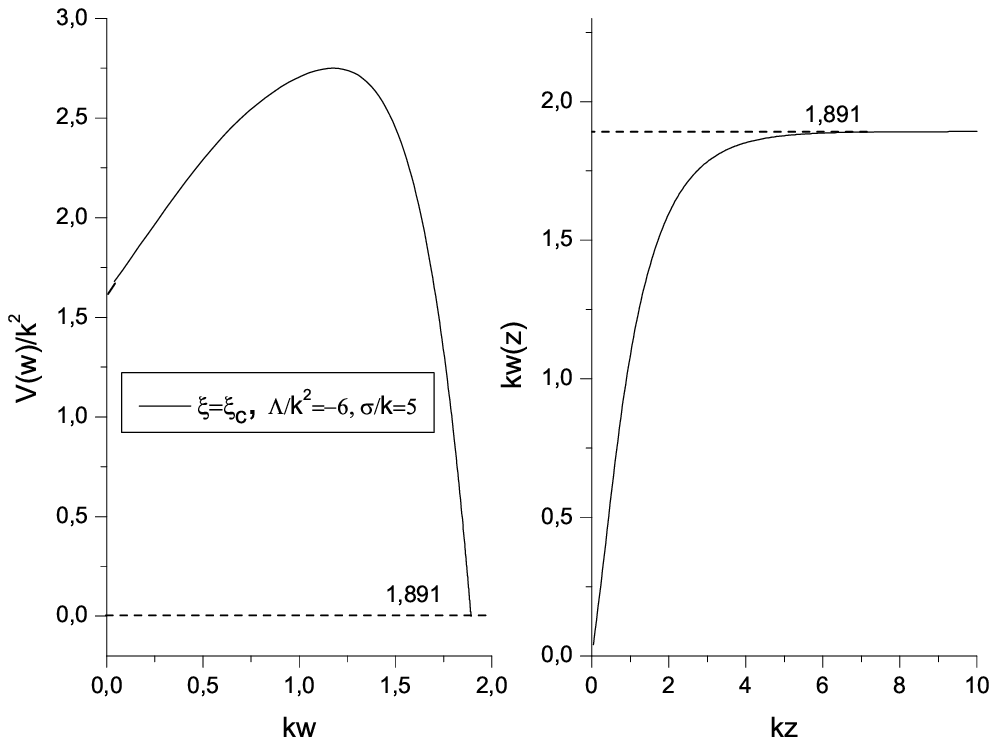}
\end{center}
\caption {as a function of kz for $\sigma=5$, $\Lambda=-6$, $k=1$,
$\lambda=0.01$. } \label{conpot}
\end{figure}

In this appendix we show that the solutions for $\xi=\xi_c$ are
stable against scalar field perturbations. In the left-hand panel of
Fig. \ref{conpot} we have plotted the potential $V(w)$ (see Eq.
(35)) as a function of $w$. We observe that the potential is always
positive and vanishes for a finite value of the coordinate $w$ ($k
w=1,891$). As we see in the left-hand panel of Fig. \ref{conpot}
this finite value for $w$ corresponds to infinite proper distance
$z$. Thus, from the form of the potential $V(w)$ in Fig.
\ref{conpot} we conclude that the scalar field spectrum consists of
continuous modes with positive energies, which becomes discrete in
the case of a second brane in the bulk. This result implies the
stability of the solutions for $\xi=\xi_c$.

\end{document}